\documentstyle[12pt]{article}
\begin{document}
\title{Phase diagrams in the Blume-Emery-Griffiths Ising thin films}
\author{H. Ez-Zahraouy \thanks{e-mail: ezahamid@fsr.ac.ma}, 
L. Bahmad and A. Benyoussef
 \vspace{1 cm}\\
Facult\'e des Sciences, D\'epartement de Physique,  \\
Laboratoire de Magn\'etisme et Physique des Hautes Energies.\\
{\small  B.P. 1014, Rabat, Morocco}}

\date{}
\maketitle
\\

\begin{abstract}
We study the spin-1 Ising model with bilinear and biquadratic exchange interactions
and single-ion crystal field. In addition to the four usual phases:
disordered DIS, ferromagnetic FER, antiquadrupolar AQU and
ferrimagnetic FRI, we found three new phases in the case of a thin magnetic film, namely: the sublattice A
non magnetic phase  NMA, the sublattice B non magnetic phase NMB and the global
non magnetic phase NMG. These phases are studied, for each layer of the film, either 
in the temperature-crystal field plane $(T,\Delta)$ or in the biquadratic coupling-crystal 
field plane $(d,\Delta)$, for different film thicknesses. 
It is found that the ferrimagnetic and sublattice phases are absent for a monolayer film. These phases appear for increasing film
thicknessses $N \ge 2$. On the other hand, the thermal behaviour of the 
layer quadrupolar moments $q_{A}, q_{B}$ and layer magnetisations 
$m_{A}, m_{B}$ are investigated for negative values of the 
biquadratic coupling and crystal field. It is also found that for fixed values of the
biquadratic coupling,  the temperature and the crystal field, each layer of the
film can belong to a different phase. To illustrate this situation, an
example is given for $d=-3.0$, $\Delta =-3.0$ and $T=1.3$.
\end{abstract}

\newpage

\section {Introduction.}
The study of ultra-thin magnetic films, with only few atomic layers thick, 
is recently subject of modern experimental methods such as the molecular beam epithaxy technique.
Very little work has been done in the case of the spin-1 Ising model also called the Blume-Emery-Griffiths BEG. This model was originally introduced \cite{BEG71} in order to explain the phase separation and
superfluidity in the $^3$He-$^4$He mixtures, as well as to describe
other multi-component physical systems, such as metamagnets,
liquid crystal mixtures, microemulsions, semiconductor allows, ... etc.
The  critical properties of this model, for positive values of both the bilinear and biquadratic couplings,
were established by different approximation techniques \cite{BEG71,MFA1,MFA2,RG1}.
The antiquadrupolar phase, also called staggered quadrupollar phase, was predicted 
and investigated on the square lattice by means of mean-field
approximation (MFA) and by Monte Carlo (MC) simulations \cite{MC1}.
The transition from the antiquadrupolar phase to the ferromagnetic phase
was established on a three-dimensional cubic lattice \cite{HB1,HB2,BA}.
The global MFA analysis \cite{HB1} on this lattice showed also
a number of other remarkable features such as doubly reentrant
behaviour and a ferrimagnetic phase which appears
between the antiquadrupolar phase and the ferromagnetic phase.
Latter investigations \cite{BA,N,NB,CVM2,CVM3} mainly confirmed these
results. Using exact solution on the Bethe lattice \cite{AKH92}
 have constructed the full set of phase diagrams for spin-1 Ising model for both positive and
negative biquadratic coupling. The diagrams showed doubly-reentrant behaviour,
staggered quadrupolar and ferrimagnetic phases, but the phase diagrams change when 
changing the coordination number.
Extending the biquadratic coupling constant to negative values \cite {HOS91}, detailed phase diagrams of the bulk BEG model were established, but neither the sublattice non magnetic nor the global non magnetic phases were outlined.
On the other hand, the usual phases of the BEG model are found in Ref. \cite{TUC98}, in the case of negative biquadratic coupling,
using the cluster variational theory in the pair approximation but only for a bilayer film with five-fold coordination. \\
Our aim in this paper is to study the BEG  spin-1 Ising film with negative values of the biquadratic coupling
for different film thicknesses $N$. The phase diagrams we obtained contain some new phases: the sublattice A
non magnetic phase  NMA, the sublattice B non magnetic phase NMB and the global
non magnetic phase NMG, in addition to the usual phases obtained by other authors, namely:
the disordered phase DIS , the ferromagnetic phase FER, 
the antiquadrupolar phase AQU and the ferrimagnetic phase FRI. 
The paper is organized as follows. In the Sect. 2 we define the model
and the used method. Resulting phase diagrams are
presented and discussed in Sect. 3. The final Sect. 4. is devoted to conclusions.

\section {Model formulation.}
The BEG model, with spins $s_i=\pm 1$, is characterized by two order parameters, magnetization $\;m\;$
and quadrupolar moment $\,q$:
\begin{equation}
\label{od}m=\langle s_i\rangle, \;\;\;\;\; q=\langle s_i^2\rangle
\end{equation}
The case we are studying is a bipartite lattice, i.e. the lattice is divided on two
sublattices A and B, such that every site belonging to A is surrounded only
by sites belonging to B and vice versa.
To account the two-sublattice structure we need
four order parameters: $m_{A,B}=\langle s_i\rangle _{A,B}$
and $q_{A,B}=\langle s_i^2\rangle _{A,B}$,
where A, B denotes sublattices. \\
The  Hamiltonian describing the model is defined by:
\begin{equation}
{\cal H}=-J\sum_{<ij>}s_is_j-d\sum_{<ij>}s_i^2s_j^2+ \sum\Delta_i s_i^2
\end{equation}
where $s_i$ takes the values $\pm1,\,0$ at each lattice site, $\langle
ij\rangle $ denotes a summation over all nearest-neighbour pairs, $J$ and $d$ are, respectively, the correspondingly
bilinear and biquadratic interaction couplings. $\Delta_i$ \ is the crystal field applied on each site $'i'$
of the layer $k$ of the film formed with $N$ layers, so that:\\
\begin{equation}
\Delta_i=\Delta_k=\Delta /k^\alpha
\end{equation}
In all the following of this work we will be limited to the case $\alpha=1$. \\
The parameters defining the different phases of the BEG model are:
$$
\begin{array}{llll}
\mbox {1. The ferromagnetic phase } FER: & m_A=m_B\neq 0, & q_A=q_B \\
\mbox {2. The ferrimagnetic phase } FRI: &  0\neq m_A\neq m_B\neq 0, & q_A\neq q_B \\
\mbox {3. The antiquadrupolar phase } AQU: &  m_A=m_B=0, & q_A\neq q_B \\
\mbox {4. The disordered phase } DIS: & m_A=m_B=0, & q_A=q_B  \\
\end{array}
$$
$$
\begin{array}{lllll}
\mbox {5. The sublattice A non magnetic phase } NMA: & m_A=0 & m_B\neq 0, & q_A=0 & q_B\ne 0 \\
\mbox {6. The sublattice B non magnetic phase } NMB: & m_A\ne0 & m_B= 0, & q_A\ne0 & q_B= 0 \\
\mbox {7. The global non magnetic phase } NMG: & m_A=0 & m_B= 0, & q_A=0 & q_B= 0 \\
\end{array}
$$

\section{Phase diagrams.}
In order to investigate the existing phases of a BEG Ising monolayer film at $T=0$, we study its corresponding ground state phase diagram. This is done in Fig. $1a$ for a single layer film, $N=1$, in the biquadratic coupling-crystal field plane $(d,\Delta)$. This shows that the only existing phases at $T=0$ are only three phases: the ferrimagnetic phase FER, the sublattice B non magnetic phase NMB and the global non magnetic phase NMG. 
The NMB phase is present only for negative values of both $d$ and $\Delta$, whereas the NMG phase is found only for positive values of $\Delta$ which minimises the corresponding energy term  in the above Hamiltonian. \\
The effect of increasing temperature is shown in Fig. $1b$ representing the phase diagram in  $(T,\Delta)$ plane for a fixed biquadratic coupling $d=-1.5$. Indeed, the disordered  phase DIS is reached at higher temperatures terminating respectively, the phase
FER for negative and large values of the crystal field $\Delta$, the AQU phase for $\Delta \approx 0$ and the NMG phase for 
$\Delta \ge 1$. In addition the phase DIS appears also at very low temperatures for positive values of the crystal field $\Delta \approx 0.5$ while the AQU phase replaces the NMB phase at very low temperature values. \\
The phase evolution of the system in  the $(d,\Delta)$ plane at a fixed but non temperature  $T=1.5$, see Fig. $1c$, show that  
the DIS phase is found only for  positive values of the crystal field and terminates by the NMG phases for large values of $\Delta$. 
For negative values of the crystal field, the AQU phase transits to the FER phase and this phase persists when increasing values of $d$ even for positive values of the crystal field.  \\
It worth to note that the phases FRI and NMA are not found for a monolayer film. 
Hence we investigate the phase evolutions in a BEG bilayer film. As it is shown in Fig. $2a$, for a biquadratic coupling $d=-1.5$, the first layer k=1 phase diagram, plotted in  $(T,\Delta)$ plane, exhibits the ferrimagnetic phase FRI absent for a monolayer film. This is found only for low temperatures between the FER and AQU phases. The other phases, DIS and NMG, found for a monolayer film are still present.
The DIS phase appears for $T \ge 0.5$ and disappears for lower temperatures where the NMG phase takes place for positive values of the crystal field. In addition the AQU phase replaces the NMB phase existing at very low temperatures. The disordered phase DIS appears also at very low temperatures for positive values of the crystal field when $\Delta \approx 0.5$. Concerning the second layer of this bilayer film, Fig. $2b$ in  $(T,\Delta)$ plane, shows that in comparison with the first layer, there is two important points: 
(i) The FRI phase is still exist and its region is greater than that one found for the first layer and (ii) the DIS phase is found even for $T \le 0.5$ provided that the crystal field is kept positive. The other phases: FER and AQU and NMG are also present for this layer. \\
In order to envisage the film thickness effect on different phase, we consider here a thin film formed with $N=5$ layers. We keep the biquadratic coupling at a constant and negative value $d=-1.5$ and study the behaviour of each layer under the temperature and crystal field effect. The first layer phase diagram plotted in  $(T,\Delta)$ plane, see Fig. $3a$, shows that not only the ferrimagnetic phase FRI is reached at higher temperatures  $T \approx 1.5$ but also that the NMG phase emerges rapidly even for small and positive values of the crystal field $\Delta$. For the second layer $k=2$, see Fig. $3b$, the NMA phase takes place between the FRI and AQU phases at low temperatures, while the FRI phase is still present and the NMG phase region decreases in comparison with that one of the first layer. Concerning the deeper layers, Fig. $3c$ for the  third layer $k=3$, Fig. $3d$ for fourth layer $k=4$ and , Fig. $3e$ for the fifth layer $k=5=N$, the FRI and the NMA phases persist and some regions of the FRI are detached from each other leaving the place to the NMA phase. These figures shows that not only the NMG phase region decreases more and more when the order of the layer increases, 
but also the region occupied by this phase decreases when increasing the order layer. Moreover the DIS phase appears even at low temperatures for the last layer $k=N$.\\
To complete this study, we investigate the thermal dependence of the magnetisations $m_A$, $m_B$ and quadrupolar moments $q_A$, $q_B$ for a BEG Ising thin film with $N=5$ layers for $d=-1.5$ and $\Delta = -3.0$. Indeed, as it is shown in Fig. $4a$ for the first layer, the three phases FRI, AQU and DIS shown in Fig. $3a$ for $\Delta = -3.0$,  are well illustrated in this figure, for increasing temperatures. The FRI phase found at very low temperature is followed by the FER phase and the DIS is reached at high temperatures.
Regarding the second layer $k=2$, Fig. $4b$ shows that the NMA phase appears at low temperature followed by the FRI phase, then the FER and DIS phases are reached, respectively. The NMA phase persists for deeper layers, $k=3,4,5$, whereas the FRI region phase decreases and terminates by the DIS phase, see Figs. $4c$, $4d$ and $4e$. For all layers of the film, the usual phases: FER, AQU, DIS and FRI are present in addition to the new phases, for specific values of temperature $T$, biquadratic coupling $d$ and crystal field $\Delta$.

\section{Conclusion}
In this work we have studied the phase evolutions in the
BEG spin-1 Ising film with biquadratic exchange interactions
and single-ion crystal field. We showed that in addition to the four usual phases:
disordered DIS, ferromagnetic FER, antiquadrupolar AQU and
ferrimagnetic FRI, some new phases are present in the case of a thin magnetic film, namely: the sublattice A
non magnetic phase  NMA, the sublattice B non magnetic phase NMB and the global
non magnetic phase NMG. These phases are studied for a monolayer, bilayer and thin film,
for each layer of the film, either in the temperature-crystal field plane  or in the biquadratic coupling-crystal 
field plane. The ground state phase diagram, established for a monolayer film, shows that at $T=0$, only three phases are present, namely:
NMA, FER and NMG. Moreover, the thermal behaviour of each layer quadrupolar moments $q_{A}, q_{B}$ and  magnetisations 
$m_{A}, m_{B}$ are investigated for fixed values of the 
biquadratic coupling and crystal field. It is also shown that for fixed values of the
biquadratic coupling $d$,  temperature $T$ and the crystal field $\Delta$, the different layers of the
film exhibit different phases. 


\newpage\

\newpage\

{\bf Figures captions:}
\begin{description}
\item[Fig. 1]  For a BEG Ising monolayer film, $N=1$: \\
(a) Ground state phase diagram in $(d,\Delta)$ plane showing only three phases: FER, NMB and NMG. 
The NMB phase is present only for negative values of both $d$ and $\Delta$, 
the NMG phase is found only for positive values of $\Delta$. \\
(b) Phase diagram in  $(T,\Delta)$ plane for $d=-1.5$, the DIS phase is reached at higher temperatures surrounding
the phases FER and NMG when increasing temperature, in addition the AQU phase replaces the NMB phase. \\
(c) Phase diagram in  $(d,\Delta)$ plane for $T=1.5$, the DIS phase is found only for  positive values of the crystal field and terminates by the NMG phases for large values of $\Delta$.  \\

\item[Fig. 2]  For a BEG Ising bilayer film $N=2$ and $d=-1.5$: \\
(a) The first layer k=1 phase diagram, in  $(T,\Delta)$ plane, shows that the ferrimagnetic phase FRI absent for a monolayer film, appears for low temperatures between the FER and AQU phases. The phases: DIS and NMG, found for a monolayer film are still present.
(b) In comparison with the first layer, the second layer k=2 phase diagram, in  $(T,\Delta)$ plane,  illustrates two important points: 
(i) the region of the ferrimagnetic phase FRI is greater than that one of the first layer and (ii) the DIS phase is found even for $T \le 0.5$ provided that the crystal field is kept positive. The other phases: FER and AQU and NMG still exist for this layer. \\

\item[Fig. 3]  For a BEG Ising thin film with $N=5$ layers and $d=-1.5$: \\
(a) The properties of the bilayer film still exist for the first layer $k=1$ of this thin film,
but its phase diagram, in  $(T,\Delta)$ plane, shows that not only the ferrimagnetic phase FRI is reached at higher temperatures  $T \approx 1.5$ but also that the NMG phase emerges rapidly even for small and positive values of the crystal field $\Delta$. \\
(b) For the second layer $k=2$, the NMA phase takes place between the FRI and AQU phases at low temperatures. \\
For deeper layers (c)  third layer $k=3$, (d) fourth layer $k=4$ and fifth layer $k=5=N$ the FRI and the NMA phases persist and some regions of the FRI are detached from each other leaving the place to the NMA phase. The NMG phase region decreases more and more when the order of the layer increases. The NMG phase region also decreases when increasing the order layer, moreover the DIS phase appears for either at low temperature for the last layer $k=N$.\\

\item[Fig. 4]  Thermal dependence of the magnetisations $m_A$, $m_B$ and quadrupolar moments $q_A$, $q_B$ 
for a BEG Ising thin film with $N=5$ layers, $d=-1.5$ and $\Delta = -3.0$. \\
(a) The three phases: FRI, AQU and DIS shown in Fig. 3a for $\Delta = -3.0$,  are well illustrated in this figure, for increasing temperatures. The FRI phase found at very low temperature is followed by the FER phase and the DIS is reached at high temperatures.
(b) The NMA phase appears at low temperature followed by the FRI phase, then the FER and DIS phases are reached, respectively.
The NMA phase persists for deeper layers whereas the FRI region phase decreases (c), (d) and (e) and terminates by the DIS phase. 

\end{description}
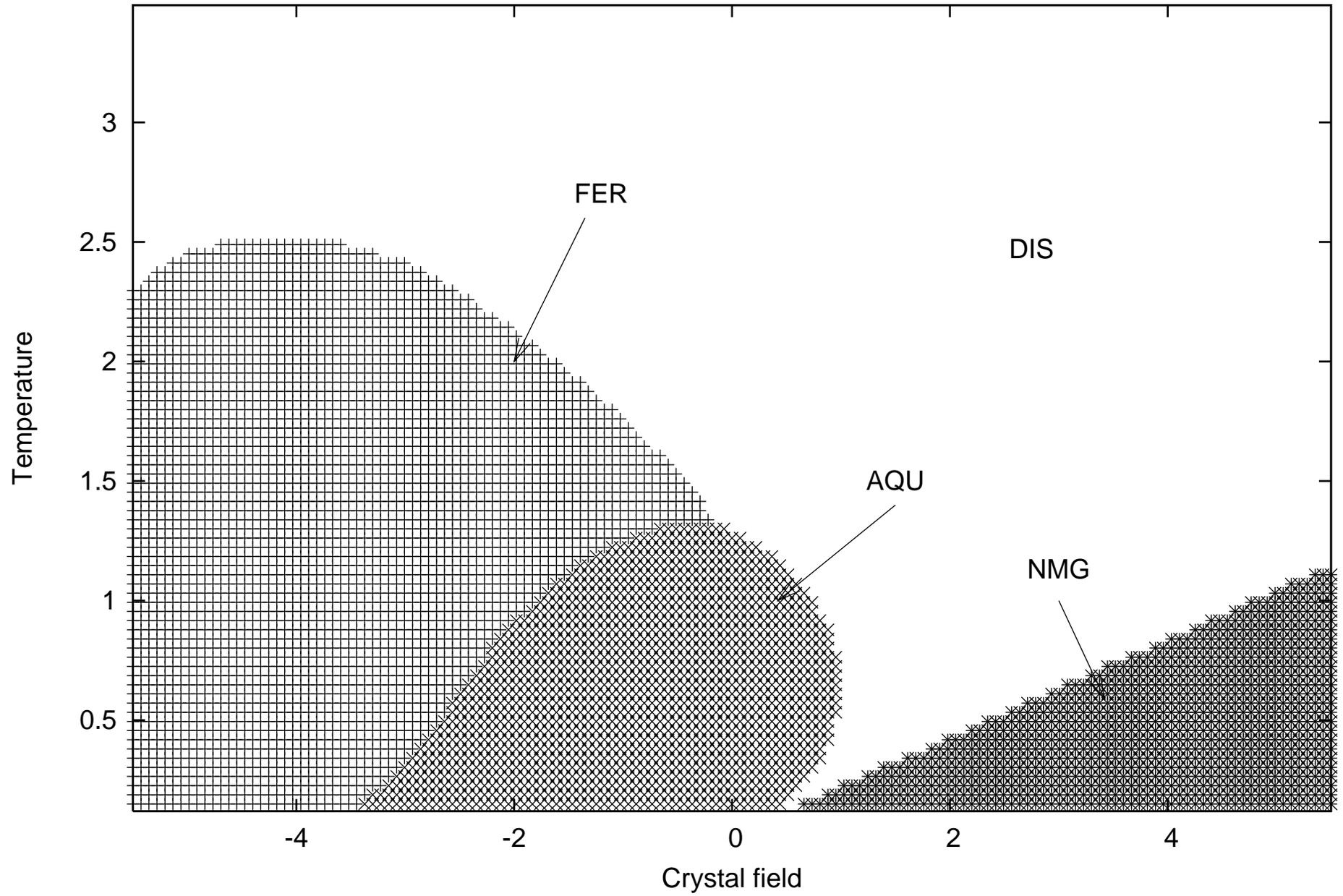
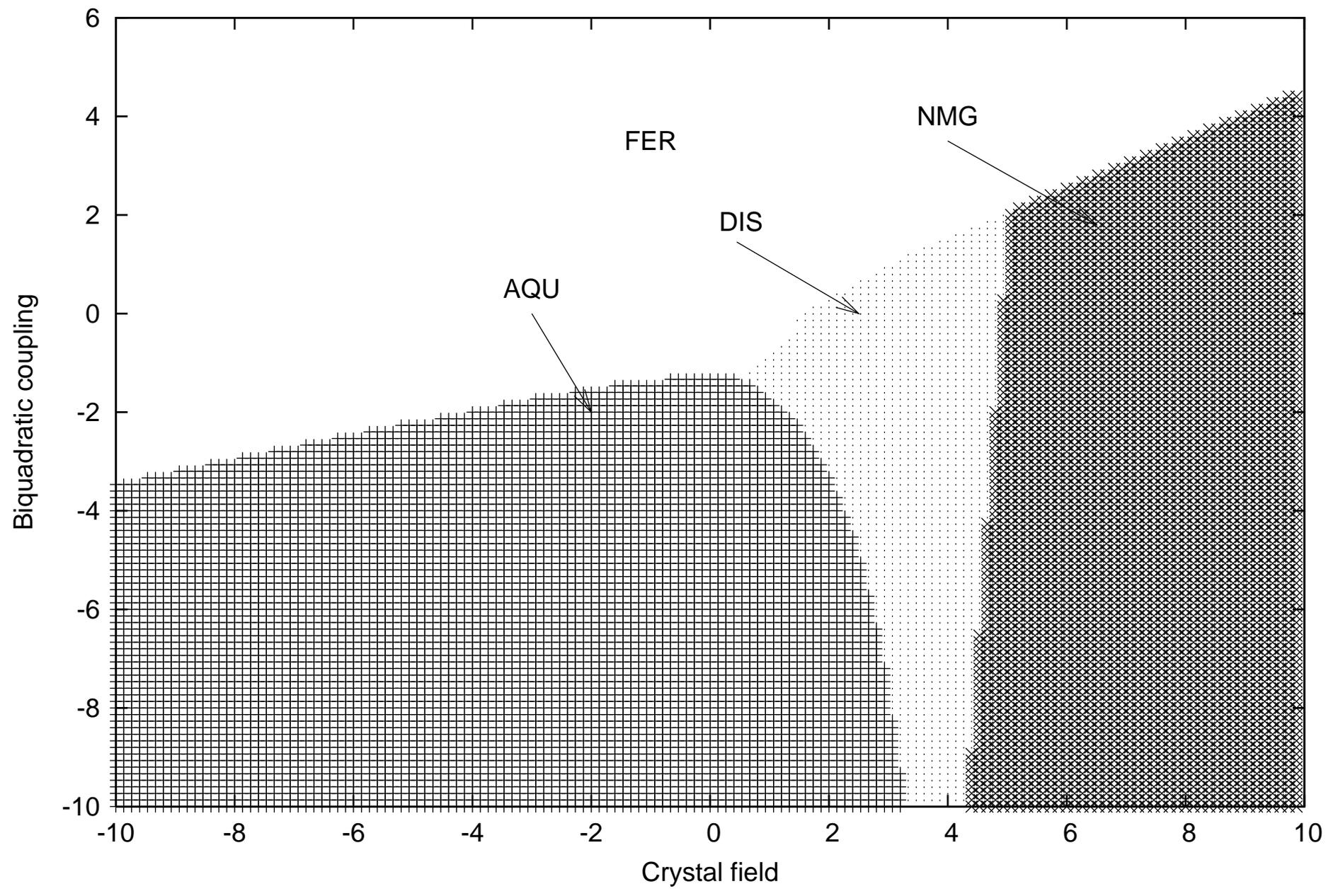
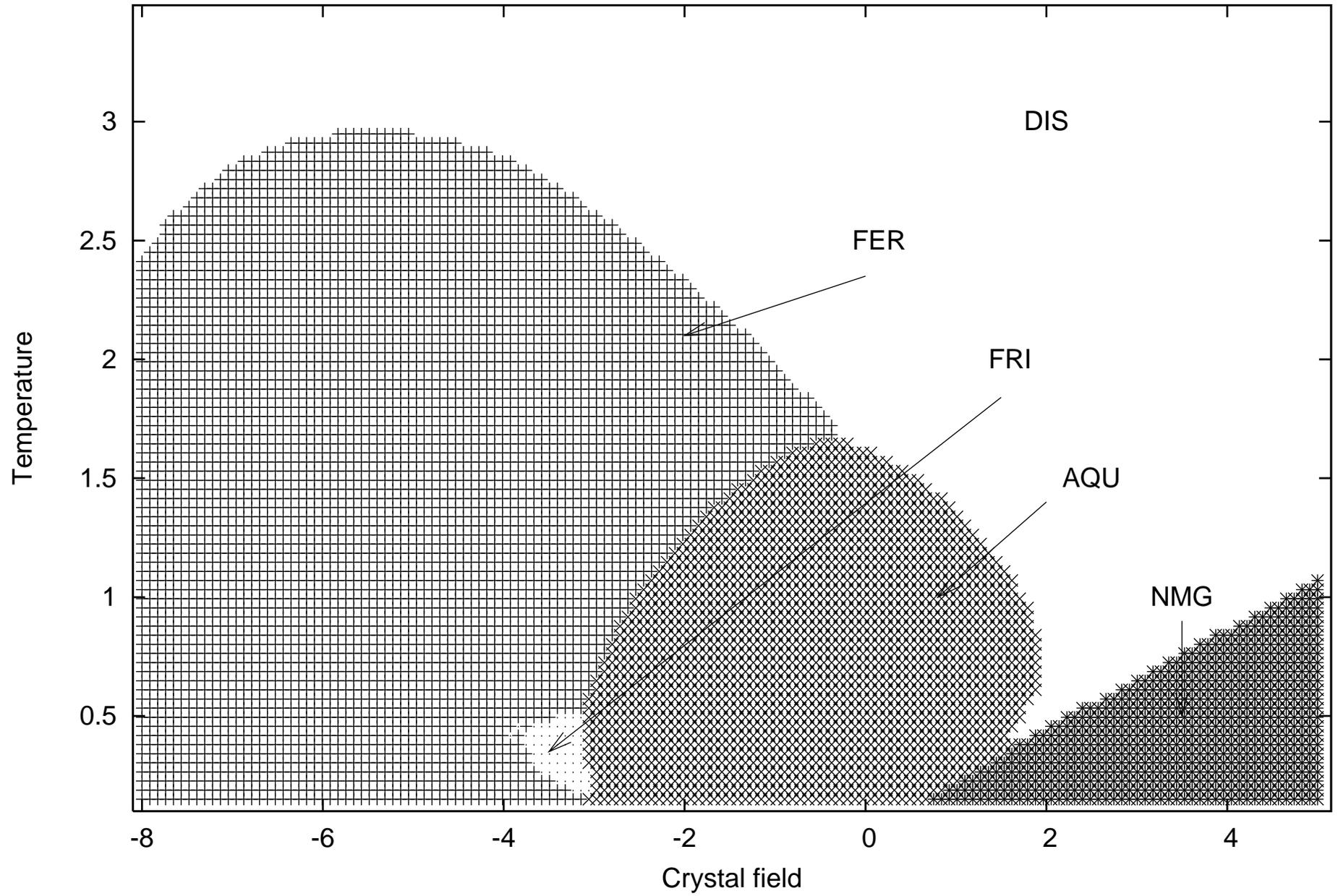
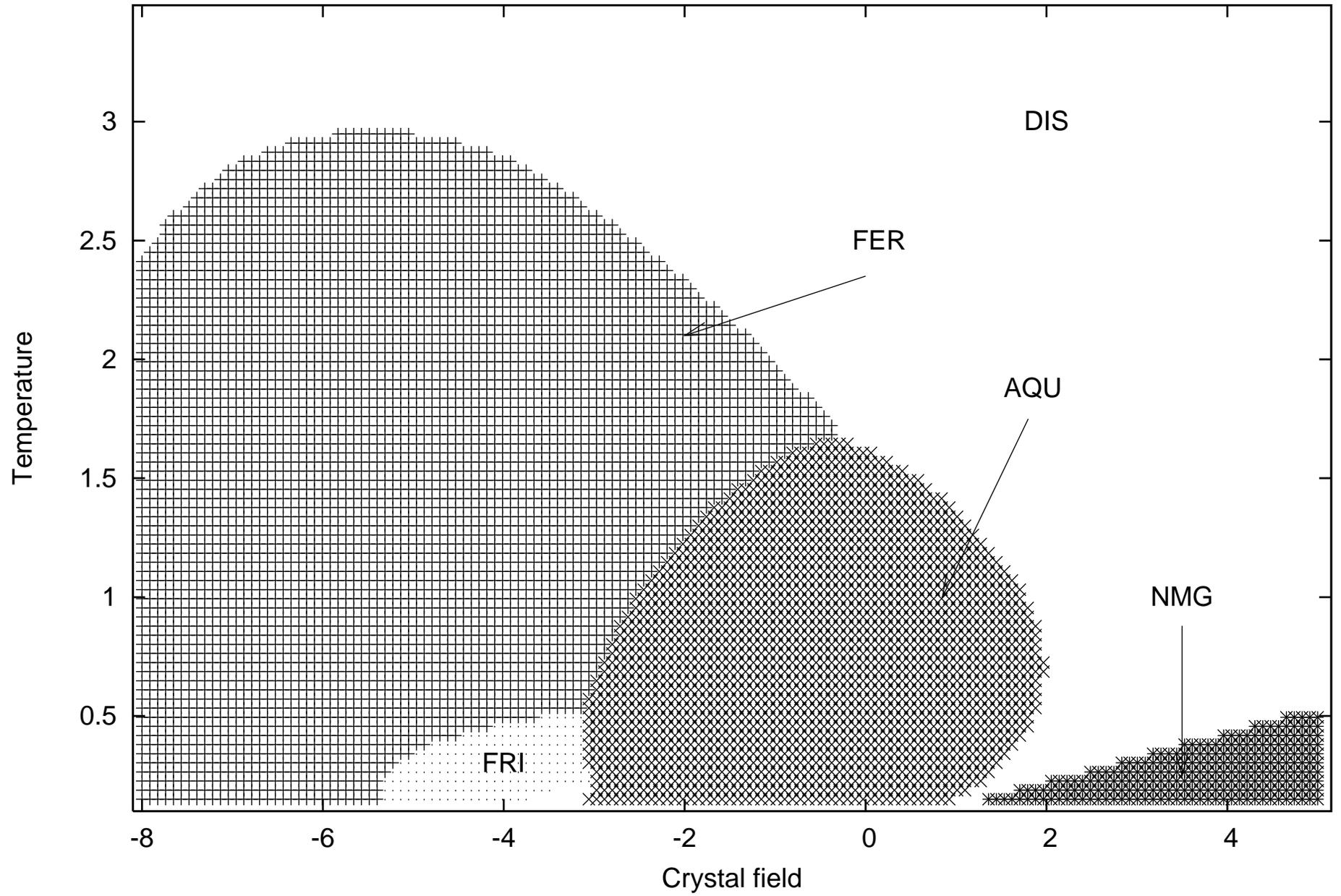
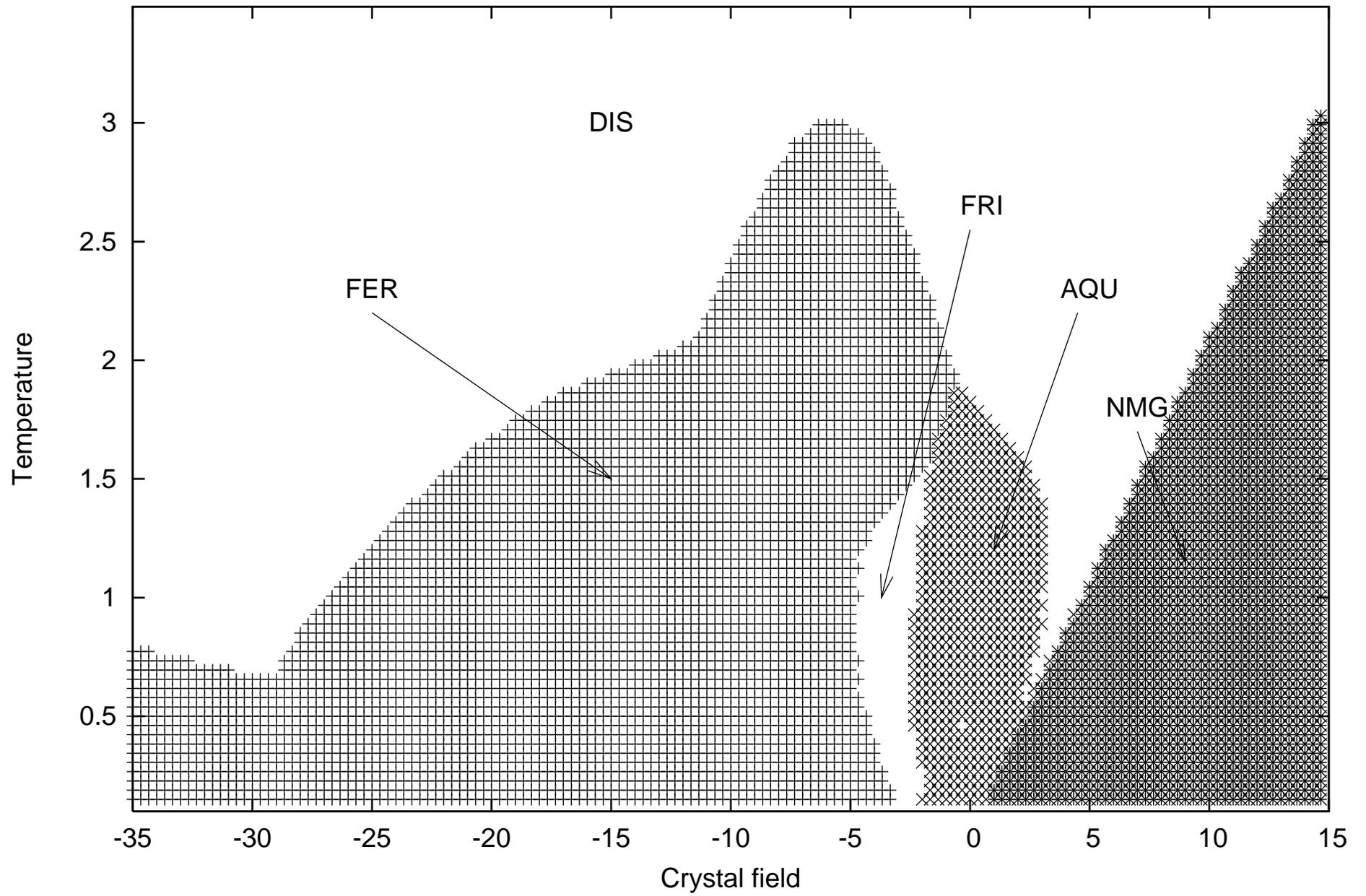
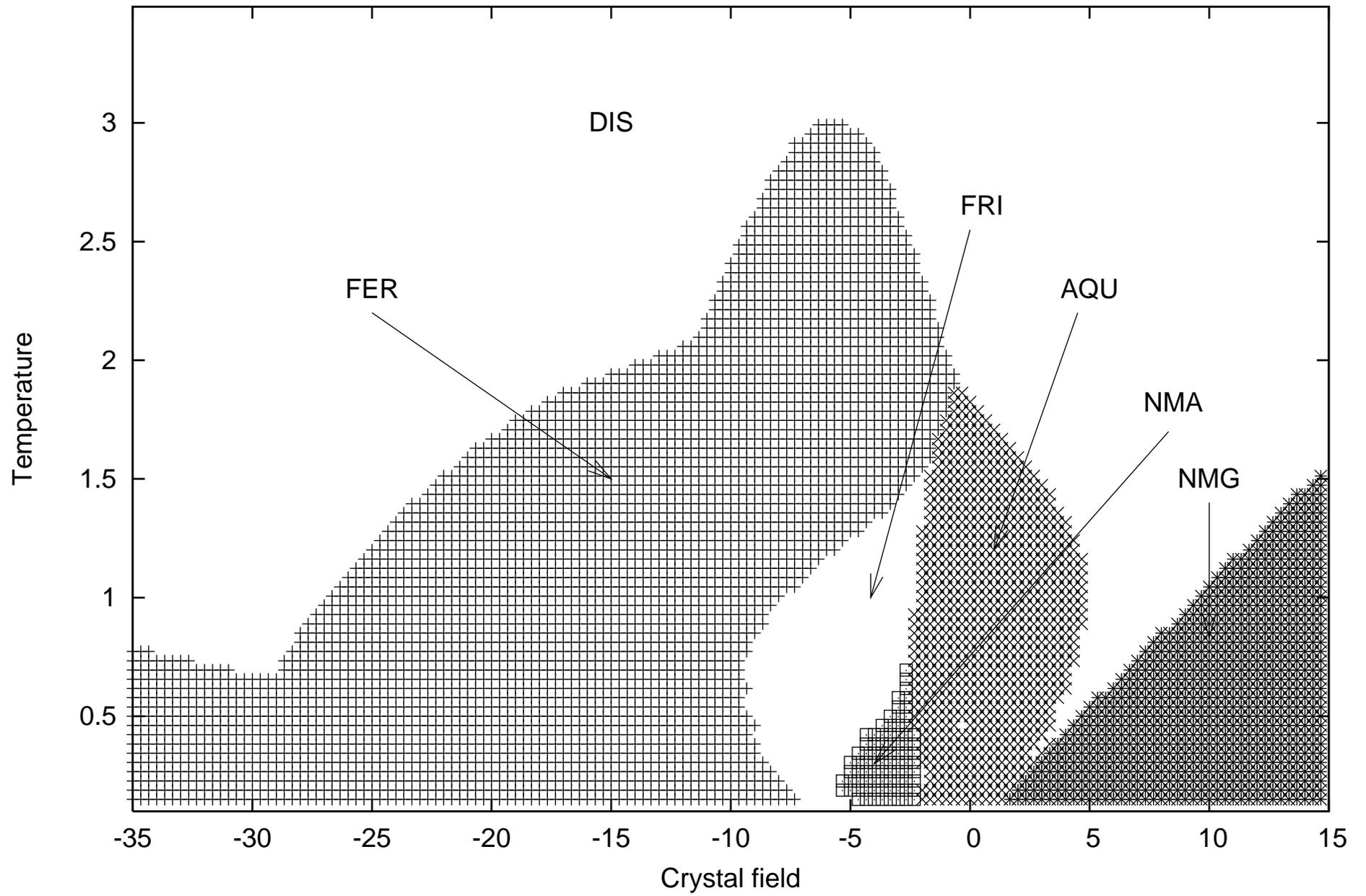
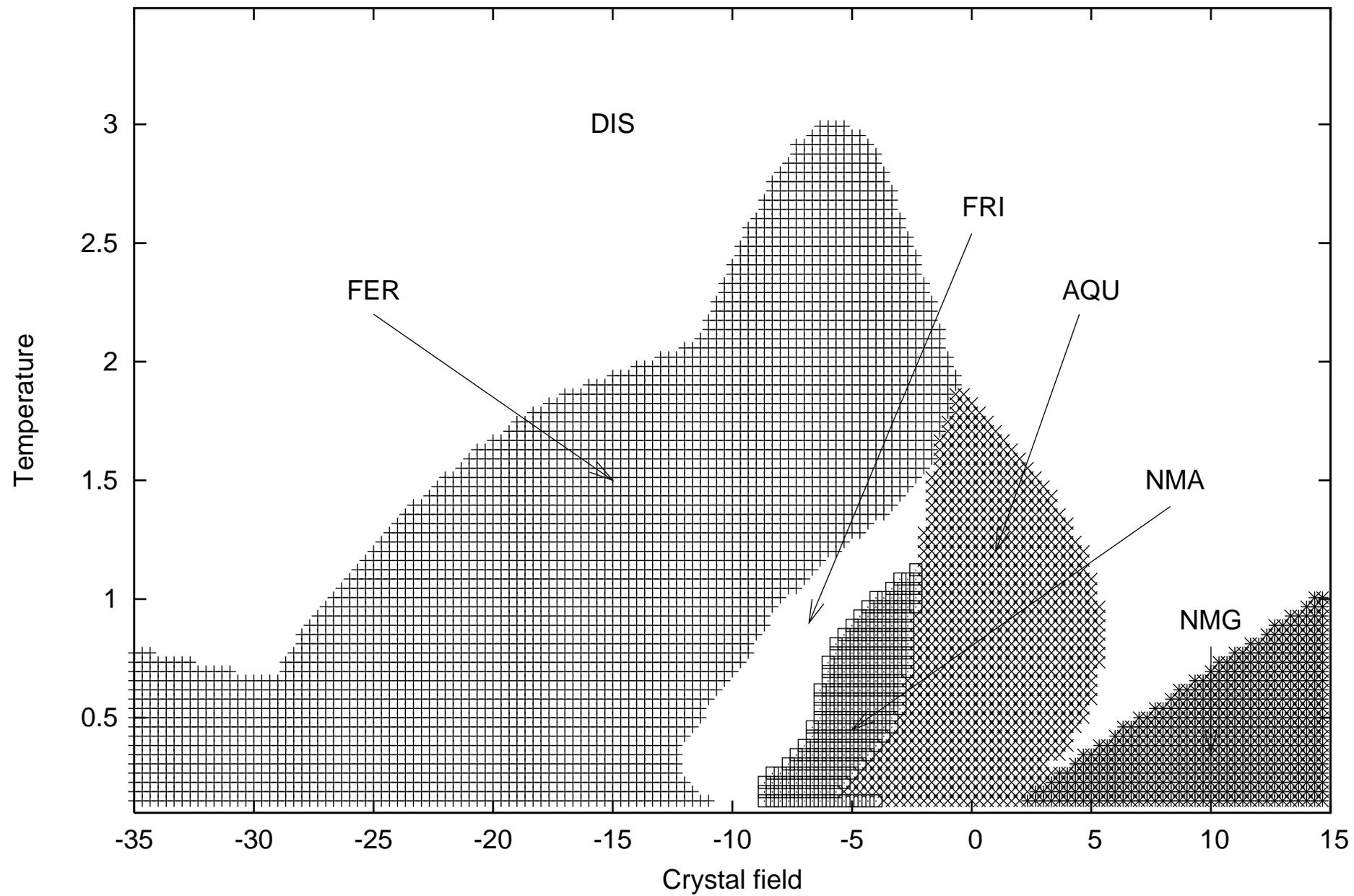
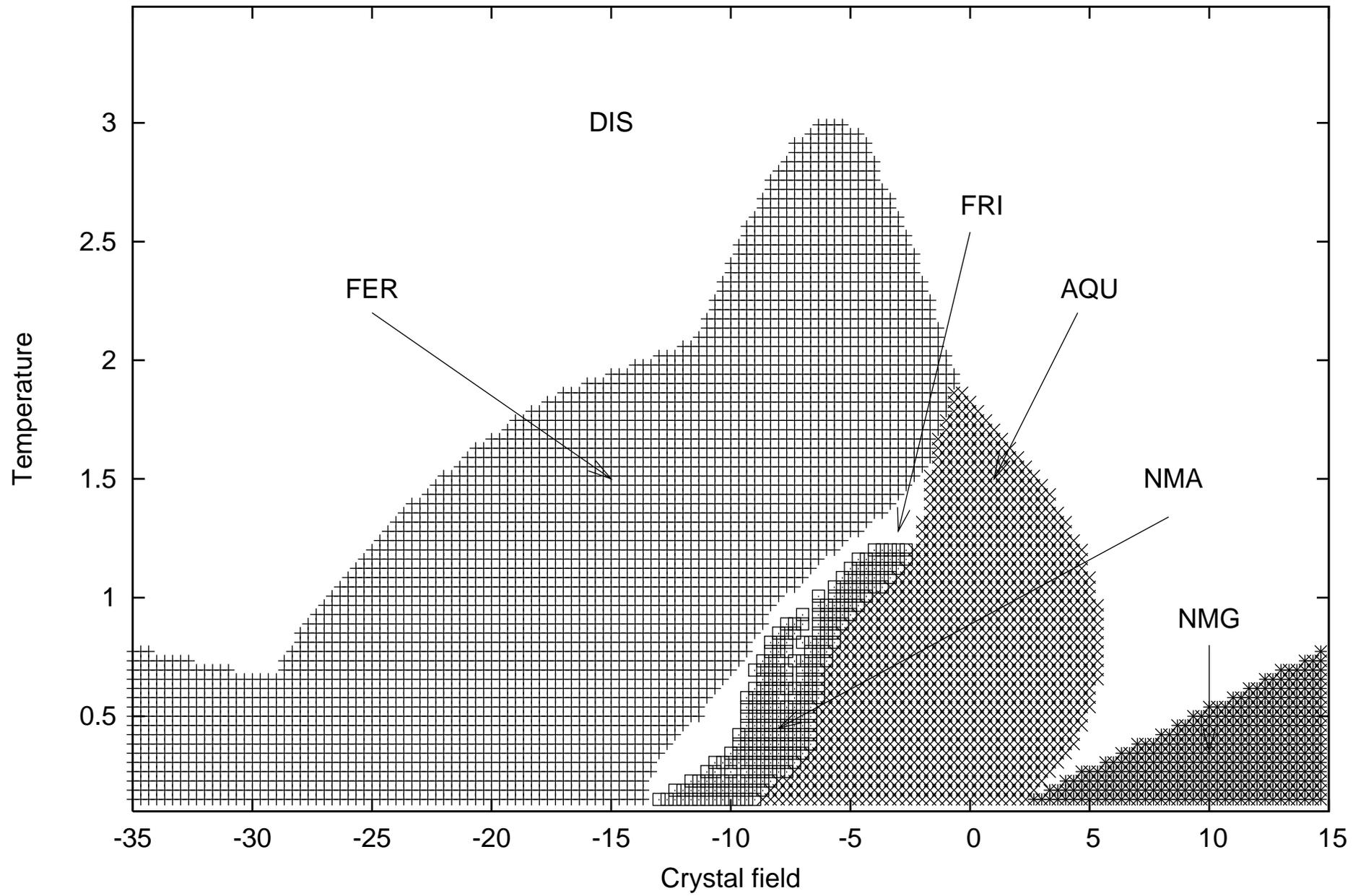
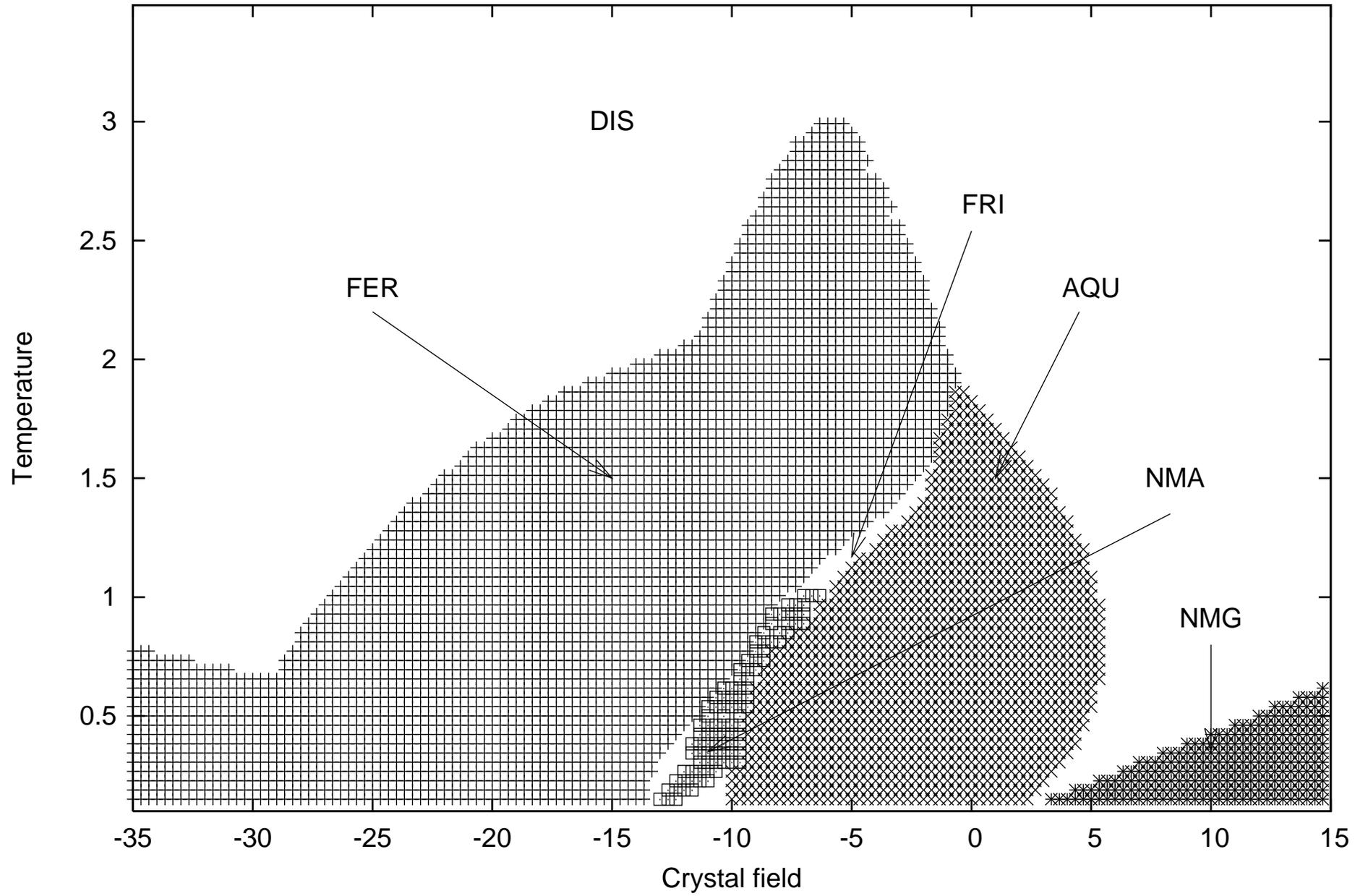
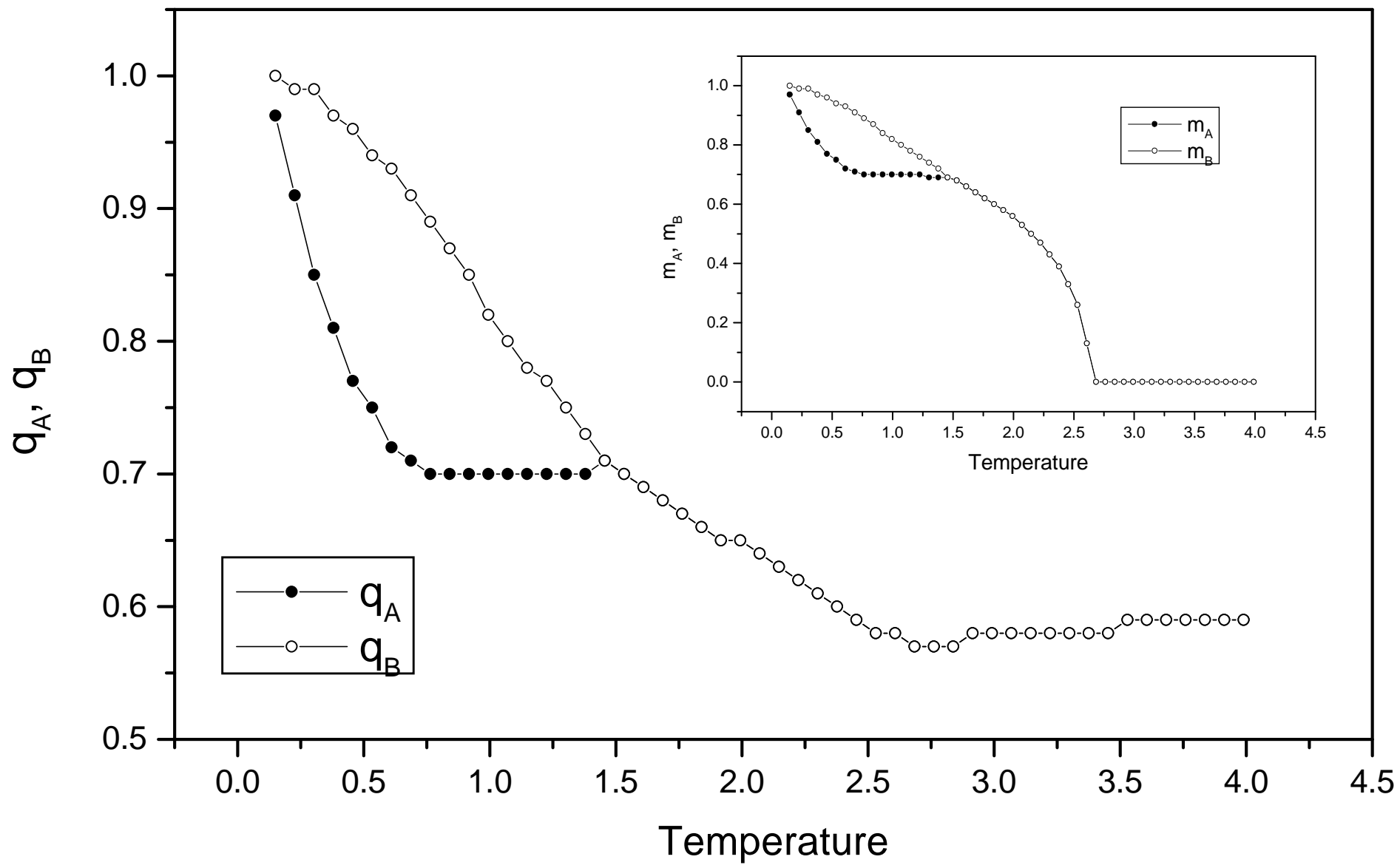
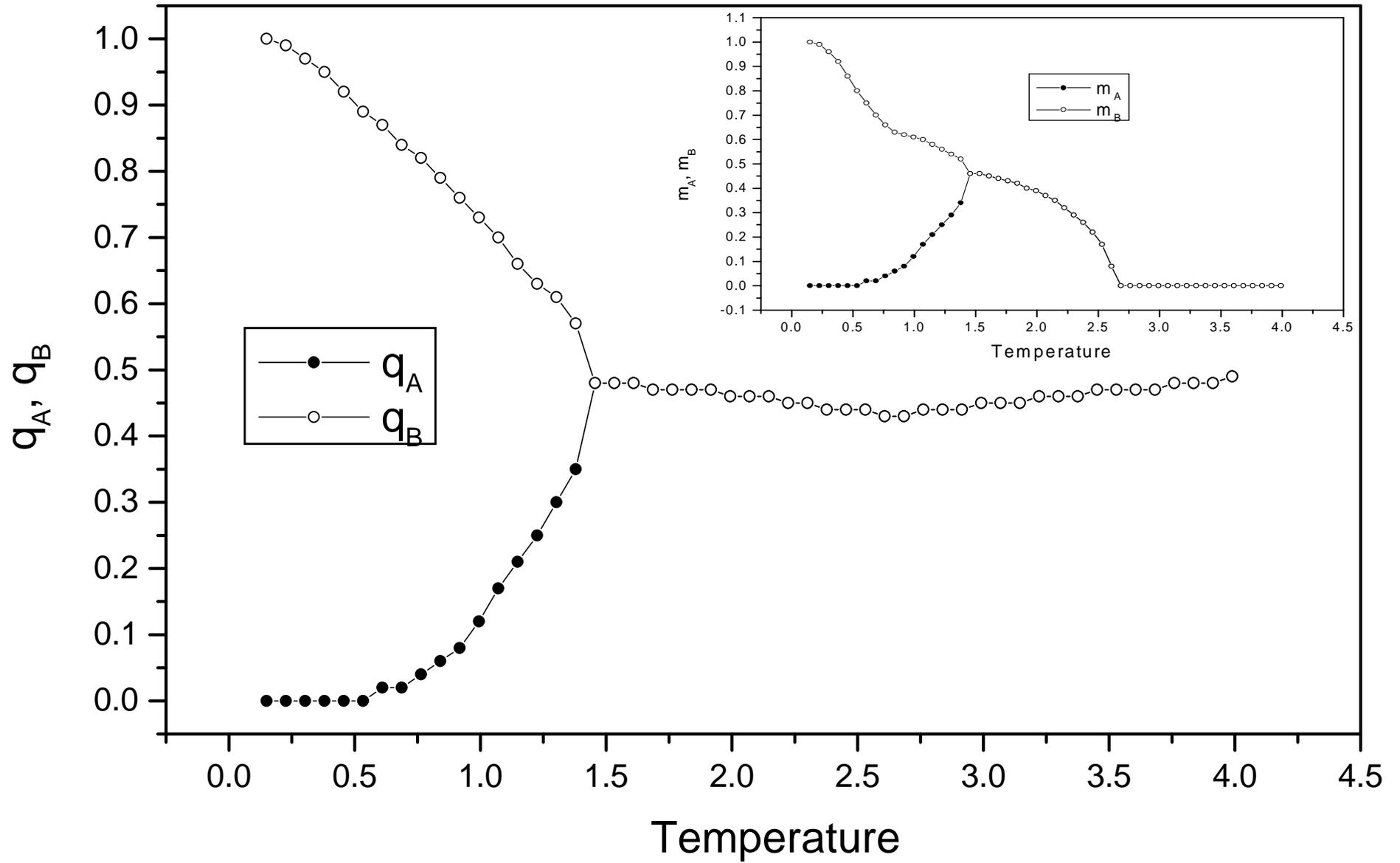
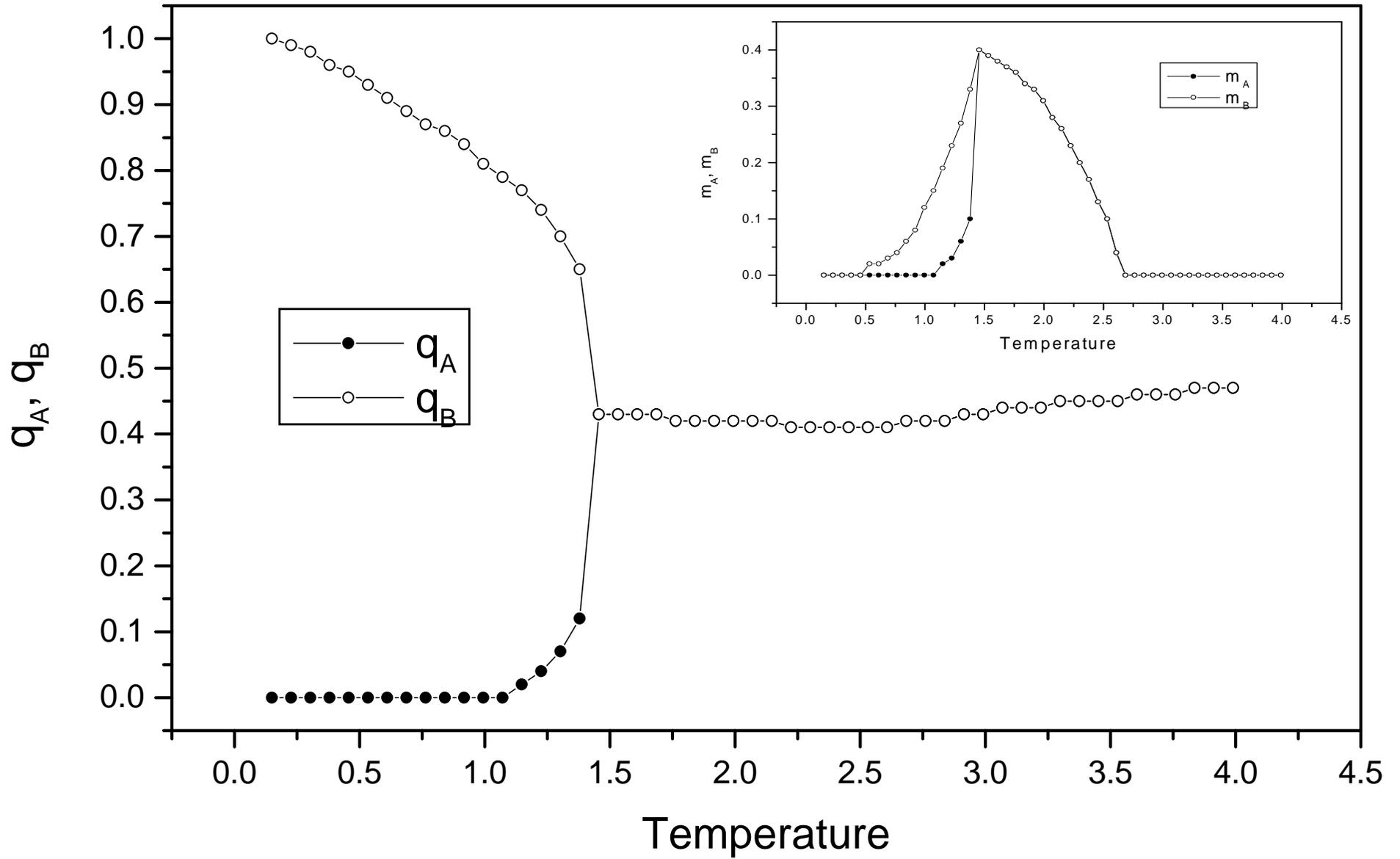
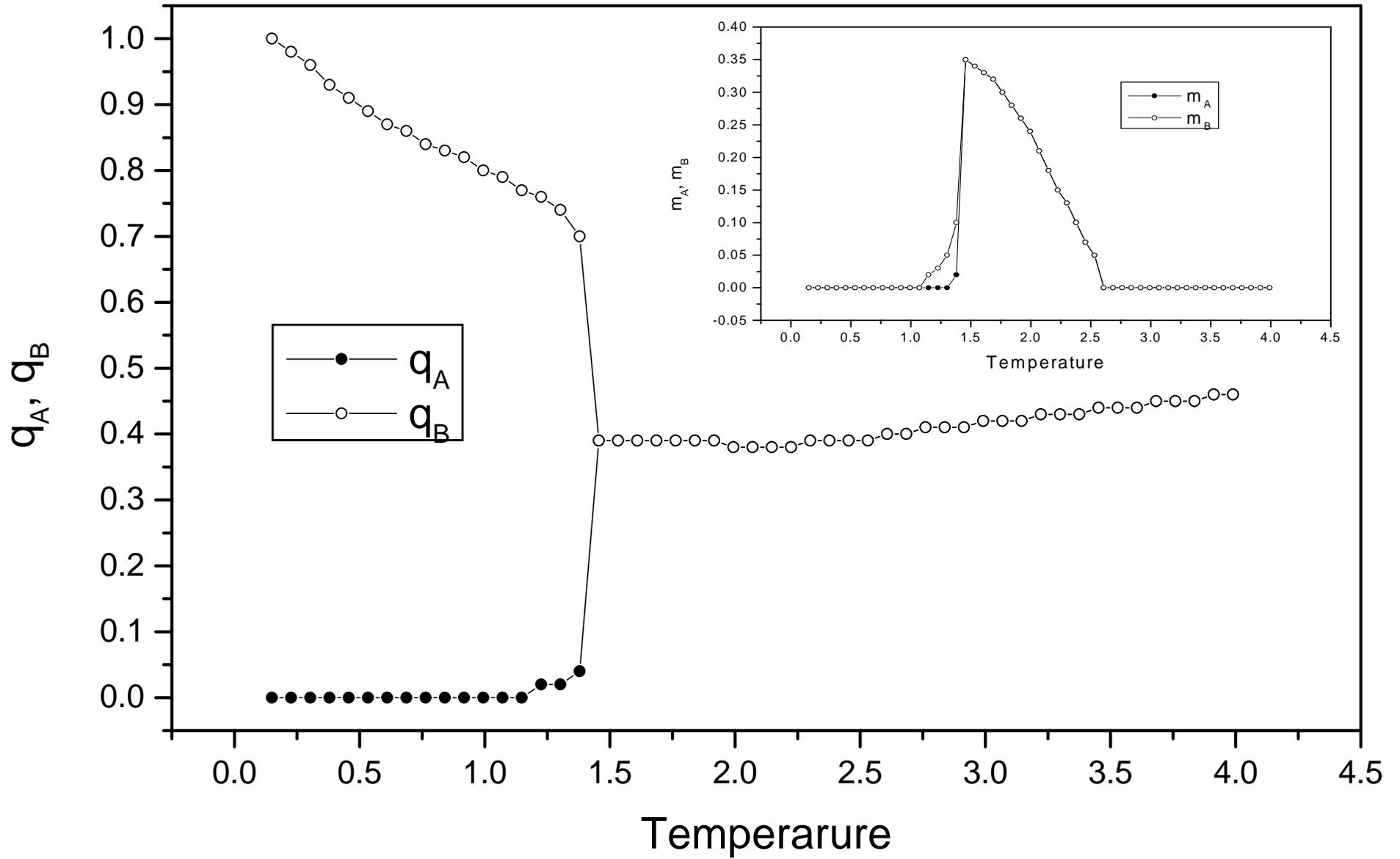
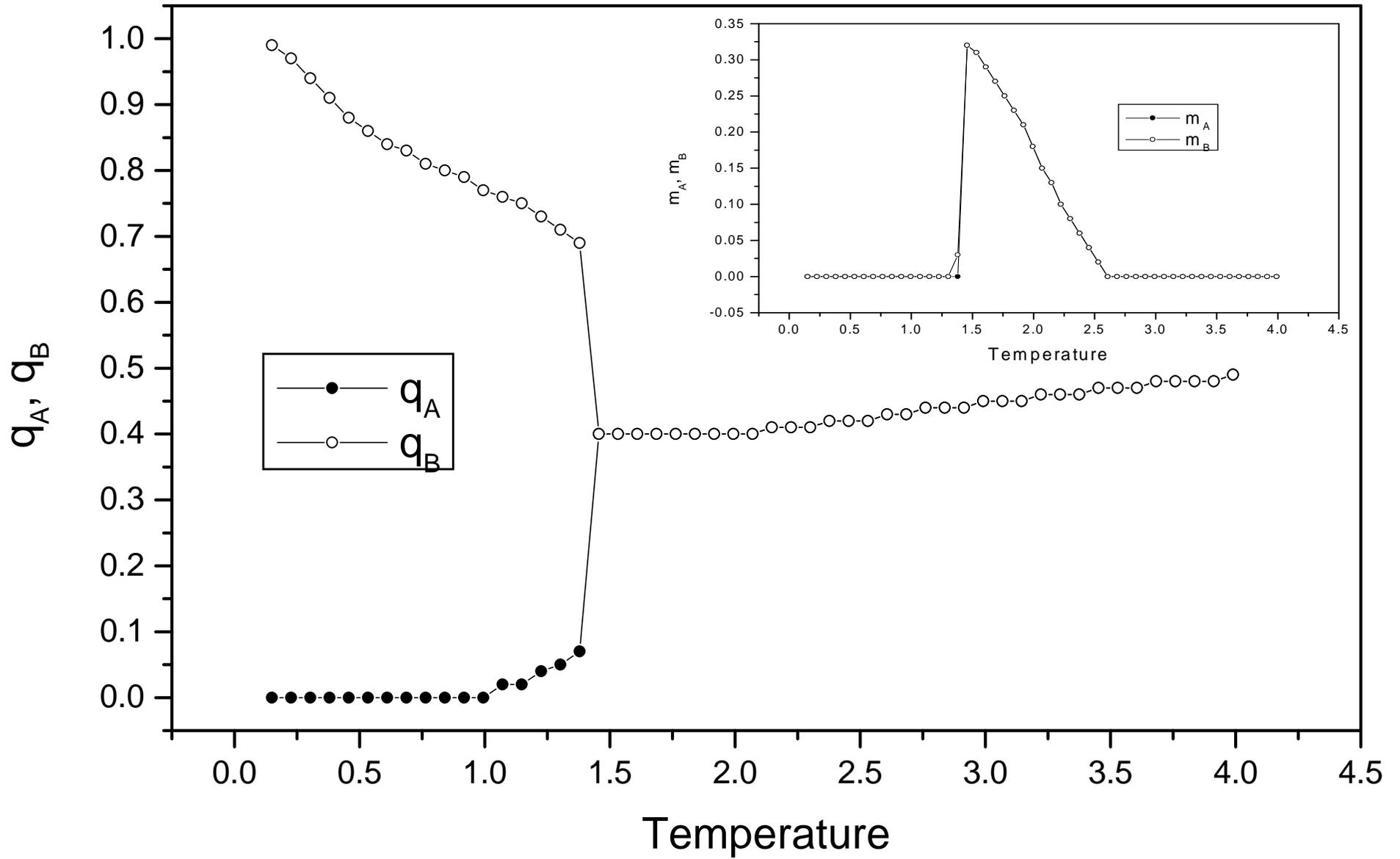
\end{document}